\documentclass[rnote]{aa} 

\usepackage{graphicx}
\usepackage{amsmath}
\usepackage{amssymb}
\usepackage{amsfonts}
\usepackage{amscd}
\usepackage{enumerate}
\usepackage{mathrsfs}
\usepackage{natbib}

\begin{document}

\title{Non-isothermal filaments in equilibrium}

\author{S. Recchi\inst{1}\thanks{simone.recchi@univie.ac.at} \and A.
  Hacar\inst{1} \and A. Palestini\inst{2}}

\offprints{S. Recchi}

\institute{Institute of Astrophysics, Vienna University,
  T\"urkenschanzstrasse 17, A-1180, Vienna, Austria \and 
  MEMOTEF, Sapienza University of Rome
Via del Castro Laurenziano 9, 00161 Rome, Italy}

\date{Received; accepted}

 
\abstract{
  The physical properties of the so-called Ostriker isothermal
  filament (Ostriker 1964) have been classically used as benchmark to
  interpret the stability of the filaments observed in nearby clouds.
  However, recent continuum studies have shown that the internal
  structure of the filaments depart from the isothermality, typically
  exhibiting radially increasing temperature gradients.
%
}{
The presence of internal temperature gradients within filaments suggests 
that the equilibrium configuration of these objects should be therefore 
revisited.  The main goal of this work is to theoretically explore how the 
equilibrium structure of a filament changes in a non-isothermal configuration.
}{
We solve the hydrostatic equilibrium equation
assuming temperature gradients similar to those derived from observations.
}{
We obtain a new set of equilibrium solutions for non-isothermal filaments 
with both linear and asymptotically constant temperature gradients. Our 
results show that, for sufficiently large internal 
temperature gradients, a non-isothermal filament could present significantly 
larger masses per unit length and shallower density profiles than the 
isothermal filament without collapsing by its own gravity.
}{
We conclude that filaments can reach an equilibrium configuration under 
non-isothermal conditions. Detailed studies of both the internal mass 
distribution and temperature gradients within filaments are then needed 
in order to judge the physical state of filaments.
}

\keywords{stars: formation -- ISM: clouds -- ISM: kinematics and dynamics -- 
ISM: structure}

\maketitle


\section{Introduction}
\label{sec:intro}


Although the observations of filaments within molecular clouds have
been reported since decades \citep[e.g.][]{se79}, only recently their
presence has been recognized as a unique characteristic of the
star-formation process.  The latest Herschel results have revealed the
direct connection between the filaments, dense cores and stars in all
kinds of environments along the Milky Way, from low-mass and nearby
clouds \citep{and10} to most distant and high-mass star-forming
regions \citep{mol10}.  As a consequence, characterizing the physical
properties of these filaments has been revealed as key to our
understanding of the origin of the stars within molecular clouds.

Classically, filaments have been interpreted assuming that $(i)$ they
are isothermal, $(ii)$ they are isolated, $(iii)$ they can be modeled
as cylindrical structures with infinite length, and $(iv)$ that their
support against gravity comes solely from thermal pressure.  However,
observational evidence is mounting that none of the above hypotheses
can be considered strictly valid:
$(i)$ recent continuum observations of filaments in different clouds
have shown that the dust temperature gradually decreases towards the
main axis of these structures \citep[e.g.][]{ste03, palm12};  $(ii)$
filaments are typically found forming intricate networks \citep[e.g.
hub-filament associations,][]{mye09} or even compact bundles of
small-scale filaments \citep{hac13}; $(iii)$ Filaments with aspect
ratios of $\sim$~4--5 are not uncommon \citep{ht11};
and $(iv)$ millimeter line studies show that the molecular emission
arising from the filaments exhibit super-thermal linewidths,
suggesting that the non-thermal motions could play a non-negligible
role in their stability \citep[e.g.][]{arzo13}.
The inclusion of any of these characteristics could drastically change
the interpretation of the physical state of the filaments.  
It is then
clear that the equilibrium properties of the filaments should be
revisited. 


In this paper we concentrate on the theoretical study of
non-isothermal filaments. The main aim of this work is to show how the
equilibrium structure of a filament changes if the hypothesis of
isothermality is relaxed.  In a companion paper (Recchi et al., in
preparation) we investigate the stability and structure of
non-isolated filaments.  In a future paper we will investigate the
effect of non-thermal pressure support within the filament.



\section{Isothermal filaments}
\label{sec:iso}
Starting from the seminal paper of \citet{cf53}, the stability of isothermal filaments has been studied by a number
of authors. 
%
\citet{stodo63} and \citet{ostri64} first demonstrated that 
the radial profile of an isothermal filament in hydrostatic equilibrium can be described by:
\begin{equation}
\rho_{eq}(r)=\rho_c\left[1+\left(\frac{r}{H}\right)^2\right]^{-2}, \mbox{ with } H=\sqrt{\frac{2c_s^2}{\pi G \rho_c}}
\label{eq:equil}
\end{equation}
\noindent
where $r$ is the radial distance from the
axis, $\rho_c$ is the central density
(i.e. the density at the axis), and $c_s$ is the isothermal
sound speed (i.e. $c_s^2=p/\rho=kT/(\mu m_H)$.
The mass per unit length (or linear mass) of this isothermal
cylinder, the so-called Ostriker filament, is given by:
\begin{equation}
  M_{lin}\equiv M_{cr} = \int_0^\infty 2\pi r \rho_{eq}(r)dr = 
  \frac{2k T}{G \mu m_H}
\label{eq:mcr}
\end{equation}
\noindent
Assuming a $\mu=2.3$ as a typical mean molecular weight in
molecular clouds and for a temperature typical for the ISM of $T=10$ K,
the linear mass of an isothermal cylinder in equilibrium is then
16.6~M$_\odot$~pc$^{-1}$.  In filaments with larger linear masses,
equilibrium between pressure and self-gravity cannot be established.
Under these conditions, self-gravity prevails and the cylinder is
destined to collapse into a spindle \citep[e.g.][]{im92}.

\section{Equilibrium solutions for non-isothermal filaments}
\label{sec:niso}

In parallel to the isothermal analysis, \citet{ostri64} also developed
the theory of stability for non-isothermal cylinders.  He considered a
generic polytropic equation of state (EOS) $P=K_n\rho^{1+1/n}$ ($n$
corresponds to $1/(\gamma-1)$, where $\gamma$ is the ratio of specific
heats).  By combining the equation of hydrostatic equilibrium
${\nabla}P=\rho {\nabla} V$ (where $V$ is the gravitational potential)
and the Poisson's equation $\nabla^2 V=-4\pi G \rho$, he found the
equation $K_n(n+1)\nabla^2 \rho^{1/n}=-4\pi G \rho$.  With the change
of variables $r=a\xi \equiv \left[\frac{(n+1)K_n}{4\pi G
    \rho_c^{1-1/n}}\right]^{1/2}\xi, \sigma=(\rho/\rho_c)^{1/n}$
(originally defined as $\theta$ by Ostriker), the resulting equation
is simply
\begin{equation}
\frac{d^2\sigma}{d\xi^2}+\frac{1}{\xi}\frac{d\sigma}{d\xi}
=-(\sigma)^n,
\label{eq:ostr}
\end{equation}
\noindent 
subject to the initial conditions $\sigma(0)=1$, $\sigma^\prime(0)=0$.
Equation~\ref{eq:ostr} has real solutions only if $n\geq -1$, or
equivalently if $\gamma\geq 0$ (but see Viala \& Horedt 1974 for the
case $n < -1$).  These solutions include the isothermal solution when
$n=+\infty$ (and $\gamma = 1$).  Under these conditions, both the
linear mass and the radius of the filament in equilibrium reach a
finite value (see Ostriker 1964 for a discussion). 

Nowadays, detailed observations of the dust emission in filaments
offer us the unique opportunity to directly estimate their radial
temperature profile (see Stepnik et al.~2003 for L1506 using PRONAOS,
Nutter et al. 2008 for TMC-1 using SCUBA, and, more recently,
Arzoumanian et al.~2011 for IC 5146 or Palmeirim et al.~2013 for B211
using Herschel).  In contrast to the temperature profiles expected for
filaments in equilibrium with polytropic EOS, the dust temperature
gradients observed in nearby molecular filaments show it to be
radially increasing \citep[e.g.][]{ste03,palm12}, probably due to the
radiation field incident on the exterior of the filament complex.  It
is important to remark that it is impossible to obtain radially
increasing temperature profiles by solving Eq. \ref{eq:ostr} for
$n>-1$.  In fact, $T\propto \rho^{1/n}$, hence $T\propto \sigma$.  For
$n>0$, both $\rho$ and $\sigma$ (hence $T$) decrease outwards.  For
$n=0$, corresponding to $\rho=Const$, the well-known solution is
$\sigma=1-\frac{1}{4}\xi^2$ (decreasing outwards).  It is easy to
verify numerically that, for $-1<n<0$, $\sigma$ and $T$ still decrease
outwards, whereas $\rho$ increases outwards because
$\rho\propto\sigma^n$.  The same result has been obtained by Viala \&
Horedt (1974; see their tables 11-13). It is also interesting to
remark that values of $\gamma<1$ are expected for polytropic filaments
in gravitational collapse, according to simulations \citep{kh98}.  As
a result, the observation of such temperature profiles has been
interpreted as a signature of instability.


In order to take into account the observationally-based result of
radially increasing temperature gradients, we must thus adopt a
different approach.  We assume the gas temperature profile $T(r)$ to
be known and we recover the corresponding equilibrium density profile.
We still use the hydrostatic equilibrium relation $\nabla P = g \rho$,
reminding however that now both $T$ and $\rho$ vary with radius (we
still assume $\mu$ to be constant).  The resulting
(integro-differential) equation is:

\begin{equation}
\frac{k}{\mu m_H}\nabla (\rho T)=\rho \cdot \left(-\frac{2 G \int_0^r 
2 \pi {\tilde r} \rho ({\tilde r}) d {\tilde r}}{r}\right),
\label{eq:ni_eq}
\end{equation}
\noindent
Here we define $\rho=\theta \rho_c$, $T=\tau T_0$ and $r=H x$, where
$\rho_c$ and $T_0$ are the density and the temperature at the filament
axis, respectively, and $H$ is given by Eq. \ref{eq:equil}, with
$T=T_0$.  The resulting normalized equation is:
\begin{equation}
\theta^\prime = -\theta \left(\frac{\tau^\prime}{\tau}+\frac{8}{\tau x}
\int_0^x s \theta ds \right),
\label{eq:ni_eq_norm}
\end{equation}
\noindent
where primes indicate differentiation with respect to $x$.  This
equation is subject to the initial condition $\theta(0)=1$ (i.e. $\rho(x=0)=\rho_c$).  This
equation can be manipulated as follows: differentiate it with respect
to $x$ and use again Eq. \ref{eq:ni_eq_norm} to get rid of the
integral.  The resulting second-order ODE is: 
\begin{equation}
  \theta^{\prime\prime}=\frac{\left(\theta^\prime\right)^2}{\theta}-
  \theta^\prime\left(\frac{\tau^\prime}{\tau}+\frac{1}{x}\right)-
  \theta\left(\frac{\tau^{\prime\prime}}{\tau}+\frac{\tau^\prime}{\tau x}
  \right)-8\frac{\theta^2}{\tau},
\label{eq:maya2}
\end{equation}
where it is easy to see that the isothermal Ostriker solution, with
$\theta_i=\left[1+x^2\right]^{-2}$ (i.e. the normalized version of Eq.
\ref{eq:equil}), satisfies the above ODE if $\tau=1$ (i.e.
$T(x)=T_0$).  This form of the equation is more convenient than the
Lane-Emden equation Eq. \ref{eq:ostr} for the problem at hand.  It is
also important to note that the above equation is subject to the
initial conditions $\theta(0)=1$, $\theta^\prime(0)=-\tau^\prime(0)$.
This initial condition is due the fact that the value of the pressure
gradient at the axis must be zero (see below).  The consequence of
this initial condition is relevant: if the temperature profiles have
positive gradients at the axis (as it appears from the observations),
the density profiles must have negative gradients, i.e. they must show
central cusps.


\begin{figure*}
\centering
\begin{minipage}[b]{0.45\linewidth}
\centering
\includegraphics[width=6.4cm,angle=270]{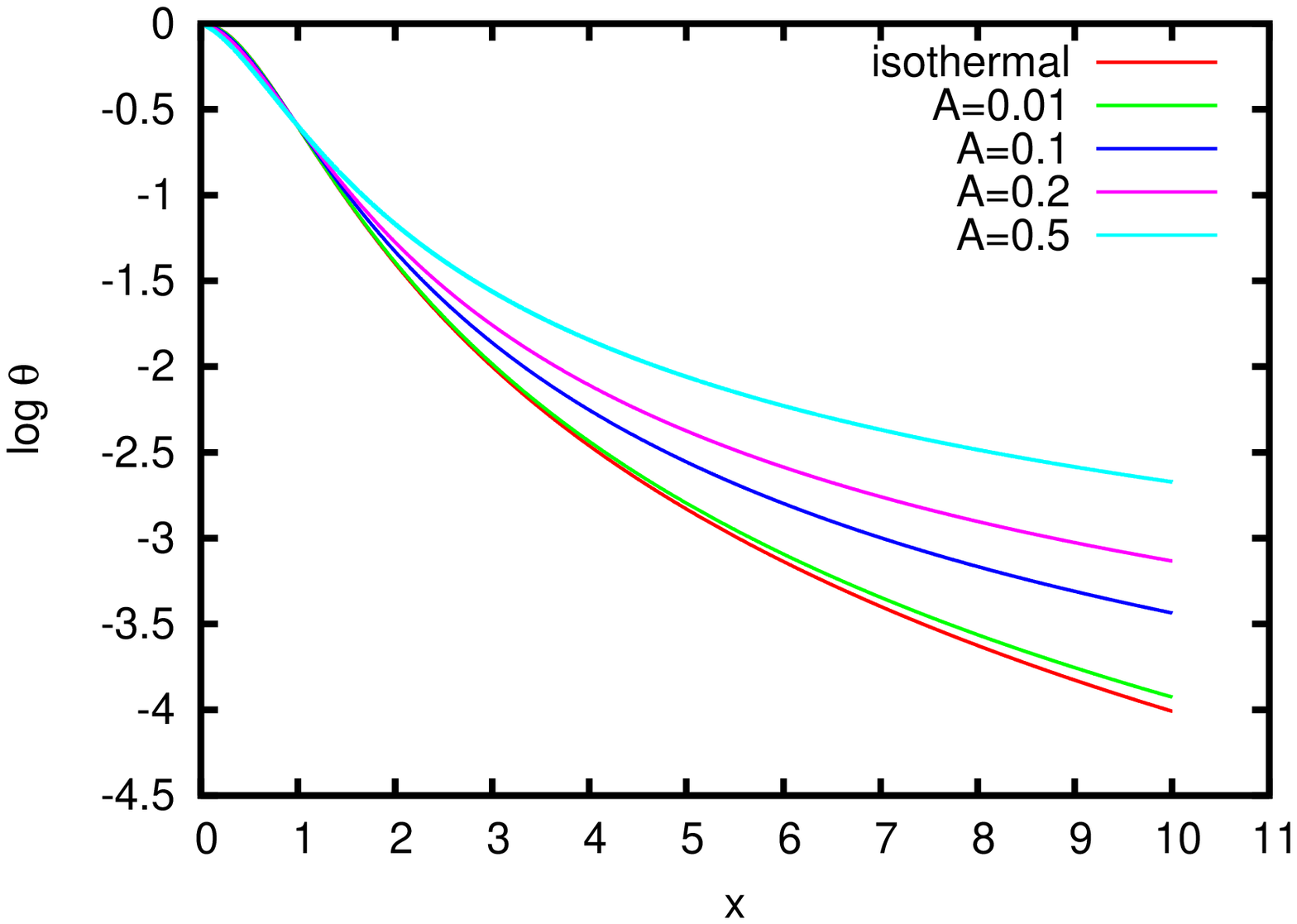}
\end{minipage}
\hspace{0.4cm}
\begin{minipage}[b]{0.45\linewidth}
\centering
\includegraphics[width=6.4cm,angle=270]{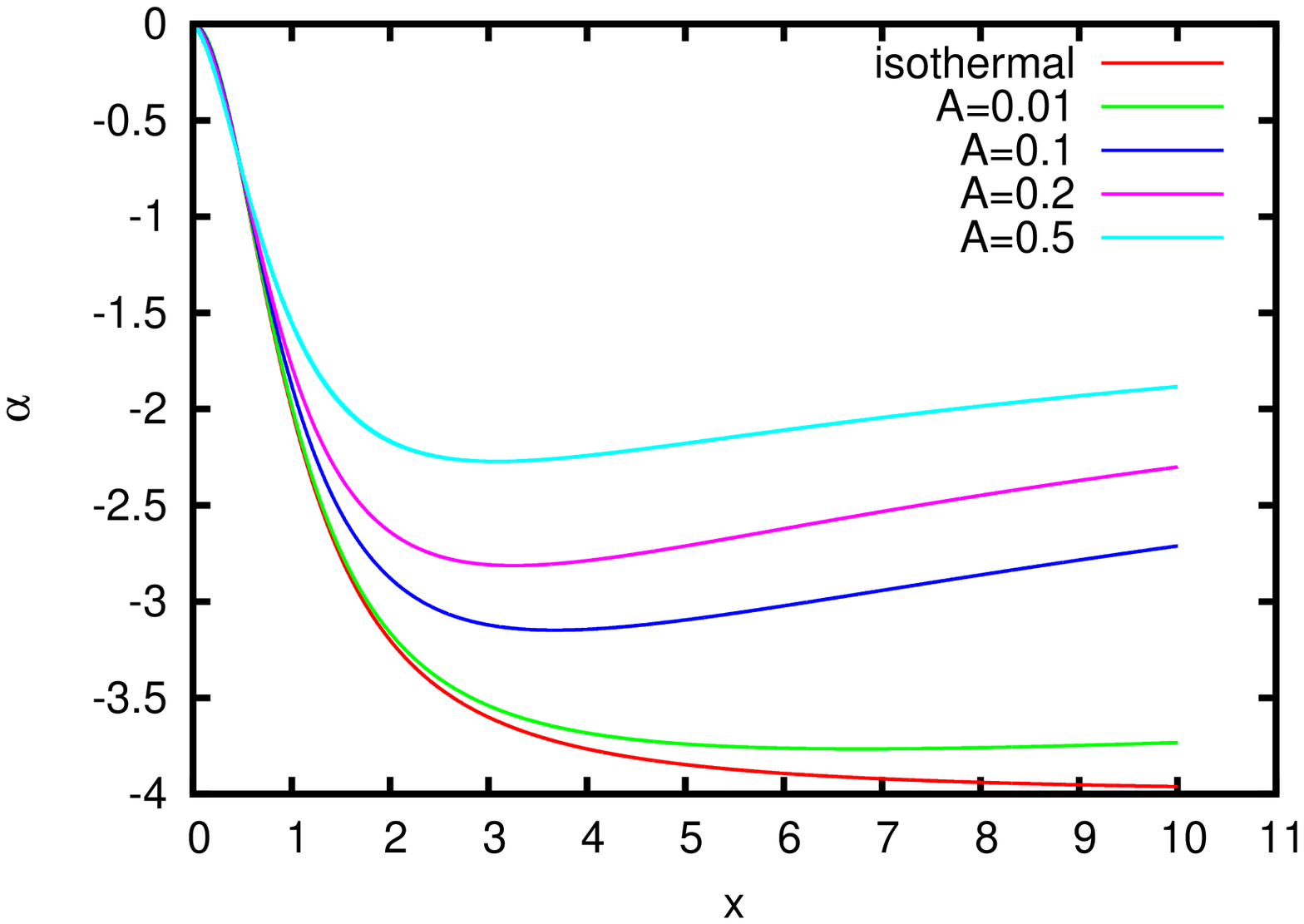}
\end{minipage}
\caption{Normalized density profiles $\theta=\rho/\rho_c$ (left panel)
  and values of $\alpha$ (right panel; see Sect. \ref{subsec:lin} for
  the definition of $\alpha$) as a function of $x=r/H$ for models with
  temperature profiles described by Eq. \ref{eq:tau1}, for different
  normalized temperature gradients $A$.  The reference isothermal
  profile is also plotted in both panels.}
\label{fig:ds1}
\end{figure*}
%
\begin{figure*}
\centering
\begin{minipage}[b]{0.45\linewidth}
\centering
\includegraphics[width=6.4cm,angle=270]{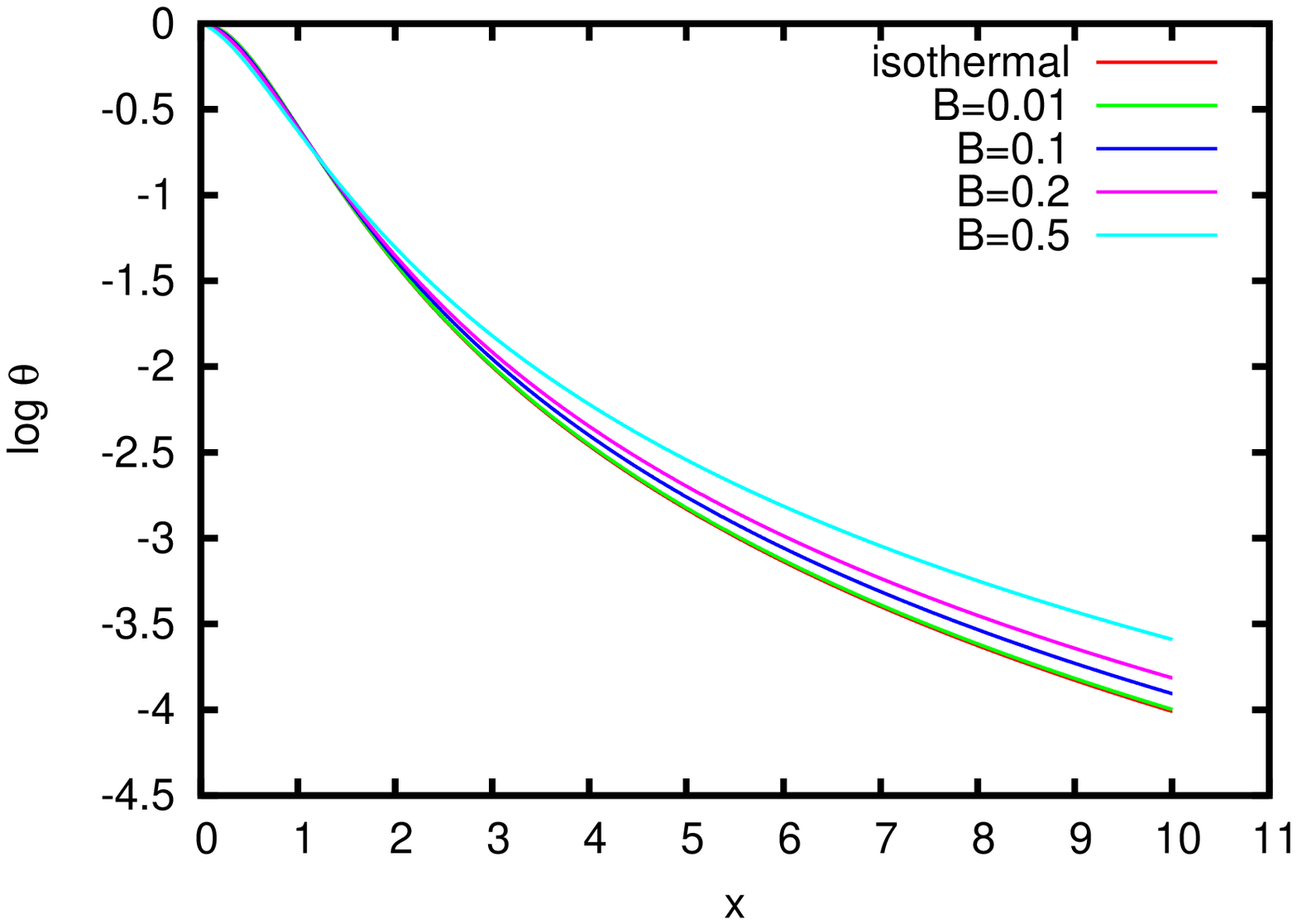}
\end{minipage}
\hspace{0.4cm}
\begin{minipage}[b]{0.45\linewidth}
\centering
\includegraphics[width=6.4cm,angle=270]{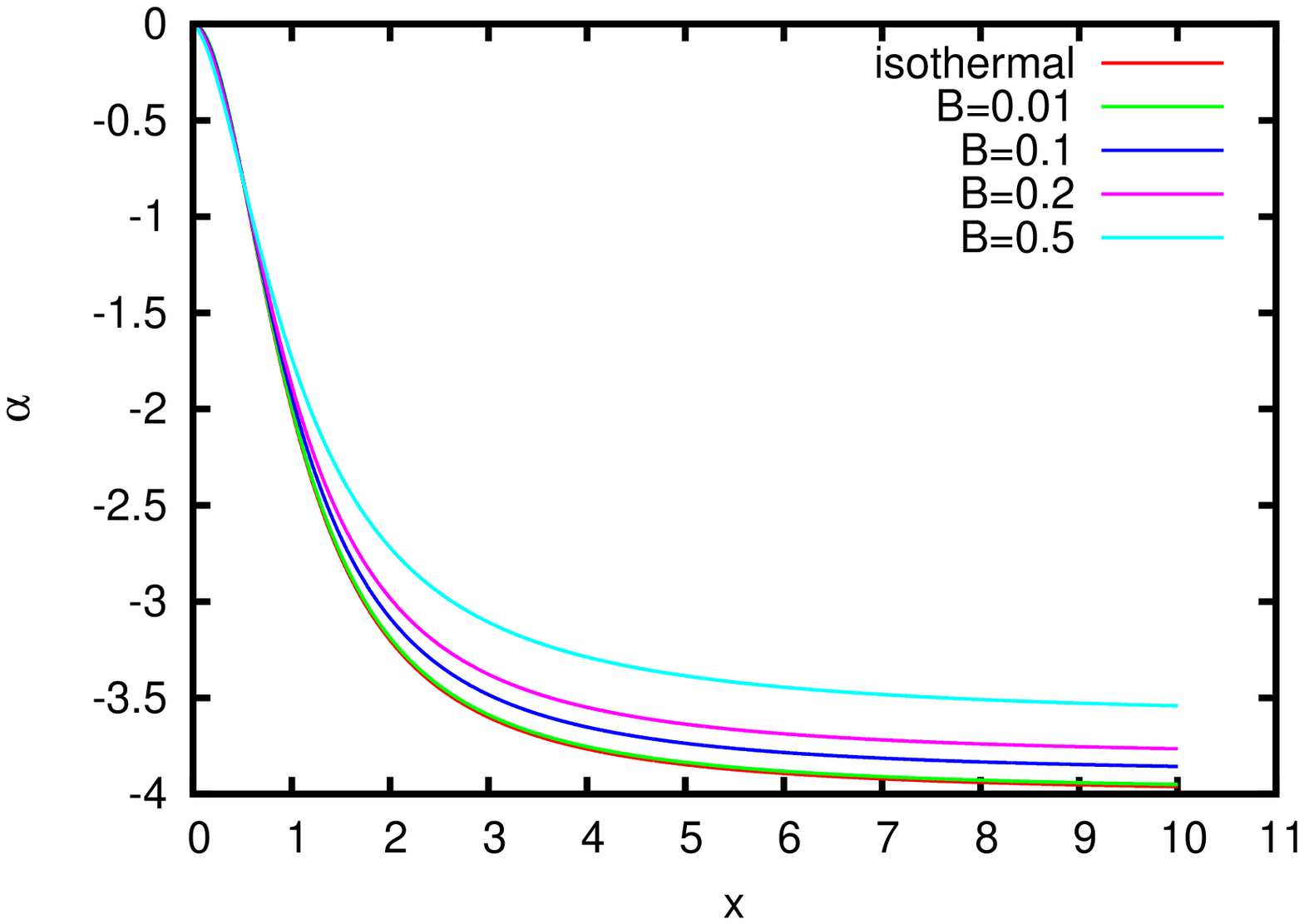}
\end{minipage}
\caption{As Fig. \ref{fig:ds1} but for models with temperature
  profiles described by Eq. \ref{eq:tau2}.}
\label{fig:ds2}
\end{figure*}

Another way to manipulate Eq. \ref{eq:ni_eq} is the following: call
$\eta$ the pressure, normalized to the central value (i.e.
$\eta=P/P_0$).  The obtained integro-differential equation for $\eta$
is $\eta^\prime=-\frac{\eta}{x \tau}\int_0^x \frac{s \eta}{\tau}ds$.
Manipulating this equation exactly as done for Eq.
\ref{eq:ni_eq_norm}, the following differential equation can be found:
\begin{equation}
\eta^{\prime\prime}=\frac{\left(\eta^\prime\right)^2}{\eta}
-\eta^\prime\cdot\left(\frac{1}{x}+\frac{\tau^\prime}{\tau}\right)
-8\left(\frac{\eta}{\tau}\right)^2.
\label{eq:maya1}
\end{equation}
\noindent
The initial conditions are $\eta(0)=1$ and $\eta^\prime(0)=0$.  This
second initial condition derives from the equation of hydrostatic
equilibrium $\nabla P = g \rho$ if we assume $g(0)=0$.  Given a
temperature profile $\tau(x)$, the density profile for a
non-isothermal filament in equilibrium $\theta (x)$ can be then
derived solving Eqs.~\ref{eq:maya2} or \ref{eq:maya1}.

\subsection{Filaments with linear temperature gradients}\label{subsec:lin}

It is instructive to consider a simplified case where the temperature
linearly increases with radius as:
\begin{equation}
\tau(x)=1+Ax,
\label{eq:tau1}
\end{equation}
\noindent
where $A$ is the normalized temperature gradient in units of the
normalized radius $x$.  This linear increase of the temperature
closely resembles the observations summarized at the beginning of this
section, at least sufficiently close to the axis.  We have numerically
solved Eq.  \ref{eq:maya2} for different values of $A$.  We have also
solved Eq.  \ref{eq:maya1} and verified that the two solutions are
identical.  This is a useful check of the consistency of our solution.
The resulting density profiles $\theta(x)$ are shown in Fig.
\ref{fig:ds1} (left panel).
%
%
We can notice from this figure that the density profile tends, as
expected, to the isothermal profile $\theta_i=\left[1+x^2\right]^{-2}$
for $A\rightarrow 0$.  Within the range of normalized temperature
gradients considered here (i.e. $0.01\leq A \leq 0.5$), the
equilibrium configuration derived from Eq.~\ref{eq:maya2} closely
resembles the Ostriker profile at short radii (at least sufficiently
close to the axis; i.e. $x<2$).  On the contrary, the derived density
profiles clearly depart from the equilibrium solution for an
isothermal filament at larger radii under the presence of temperature
gradients with $A\geq0.1$.  Indeed, if $A=0.5$ the density profile for
an non-isothermal filament in equilibrium can be more than one order
of magnitude higher than the expected value in the isothermal case at
$x\geq 6$.
%

In order to better characterize the density profile, we consider the
quantity $\alpha=\frac{d \ln \theta}{d \ln x}$.  If the density
profile can be approximated by a power law, $\alpha$ clearly
represents the exponent.  The functions $\alpha(x)$ for the different
density profiles discussed above as a function of the normalized
radius $x$ are plotted in Fig. \ref{fig:ds1} (right panel).  For the
isothermal profile $\theta_i(x)$ it is easy to see that
$\alpha=-\frac{4x^2}{1+x^2}$, with $\alpha$ tending to -4 for
$x\rightarrow \infty$.  Compared to the Ostriker solution, the
equilibrium configuration for a non-isothermal filament with $A>0$
presents therefore shallower profiles at large radii.  Particularly
for $0.1\leq A \leq 0.5$, the filaments tend to present slopes with
$-3\lesssim \alpha \lesssim -2$ for normalized radius within $4<x<10$.

The presence of density profiles shallower than the Ostriker solution
has important implications in the linear mass that can be supported
by these non-isothermal filaments.  As can be seen in
Fig.~\ref{fig:ds1}, $\alpha$ tends to increase for large values of
$x$.  Indeed, it is possible to show (see Appendix \ref{sec:a1}) that
$\alpha$ must asymptotically tend to a value $\alpha_{lim}\in]-2,-1[$.
This implies also that the normalized linear mass:
\begin{equation}
\Pi=\int_0^\infty 2 \pi x \theta(x) dx,
\label{eq:PI}
\end{equation}
\noindent
diverges.  This can be explained because the temperature tends to
infinity for $x \rightarrow \infty$, so the increasingly large thermal
pressure at large radii must be counter-balanced by an increasingly
large gravity.  
As a consequence, the linear mass for a non-isothermal filament in
equilibrium presenting a linear temperature gradient can be arbitrary
large.  In order to avoid the linear mass to diverge, a filament
described by a temperature profile as in Eq. \ref{eq:tau1} has to be
necessarily pressure-truncated at some radius.

Finally, we notice that the properties of the non-isothermal filaments
derived above resemble the equilibrium configuration found in
turbulent dominated filaments.  \citet{geh96} demonstrated that an
isothermal filament with a radially increasing turbulent pressure
presents a shallower density profile and larger linear mass than the
equivalent Ostriker-like filament at the same temperature.
%
Compared to the non-isothermal configuration, in these models the
increasing gravitational energy is balanced by a pressure generated
from the radial increase of the non-thermal motions, while in our case
the filament is always thermally supported.


\subsection{Filaments presenting asymptotically constant temperature gradients}\label{subsec:asym}

It is also useful to look for a temperature profile resembling Eq.
\ref{eq:tau1} for small values of $x$ but tending to a finite value
for $x \rightarrow \infty$.  These kinds of profiles are similar to
those found in the observations where the temperature flattens out for
large radii \citep[typically at radii $\sim$~0.4-0.5~pc; see
e.g.][]{ste03,palm12}.
 A function with such properties could be:
\begin{equation}
\tau(x)=\frac{1+(1+B)x}{1+x}.
\label{eq:tau2}
\end{equation}
\noindent
where $\tau(x)$ tends to $(1+B)$ for $x\rightarrow \infty$.  The
density profiles and the slopes $\alpha(x)$ for this temperature
profile are shown in Fig. \ref{fig:ds2}.  As can be seen in the
Figure, compared to the linear case, the density profiles obtained for
filaments described by Eq.~\ref{eq:tau2} are closer to the isothermal
profile if $B$ is small (i.e. $B\lesssim 0.2$), while they only became
significantly different to the Ostriker profile under the presence of
large gradients (i.e. with $B\sim 0.5$), at large radii.
Additionally, and for all the $B$ values, the slopes of the density
profiles decrease monotonically, with values of $\alpha<-3.5$ for
normalized radius $x>8$.

For those non-isothermal filaments in equilibrium with a temperature
structure described by Eq.~\ref{eq:tau2}, it is possible to show (see
again Appendix \ref{sec:a1}) that $\alpha$ must be smaller than $-2$.
This implies also that, in contrast to the linear case, the integral
Eq.  \ref{eq:PI} defining the linear mass is converging.  
%
%

In this sense, we have also calculated numerically the relation between both the 
linear mass $\Pi$
and the half-mass radius $x_{1/2}$ (the radius within which the mass
is $0.5 \cdot \Pi$) of a filament described by Eq.~\ref{eq:tau2} with
the equivalent values for the linear mass (which turns out to be equal
to $\pi$) and the half-mass radius (=1) for the Ostriker filament,
respectively, as a function of $B$. Our results show that these two quantities
can be well approximated by the following quadratic fits:
\begin{equation}
\Pi-\pi\simeq 0.681 B - 0.067 B^2,
\end{equation}
\noindent
\begin{equation}
x_{1/2}-1\simeq 0.441 B + 0.092 B^2.
\end{equation}
We can thus see that a non-isothermal filament (with temperature
increasing outwards but tending to a constant value) can sustain again
more mass than an isothermal Ostriker-like filament without being
gravitationally unstable. However, the differences in the linear mass
between these two models are in this case not very large (typically of
less than 20--30\%).


\section{Conclusions}
\label{sec:disc}

The steep radial profile, typically $\propto$~r$^{-4}$, and the
characteristic mass per unit length with 16.6~M$_{\odot}$~pc$^{-1}$ at
10~K, of the the isothermal Ostriker filament have been used to define
the equilibrium state of the filaments within molecular clouds.  As
shown in Sec.~\ref{sec:niso} the non-isothermal nature of the
filaments introduce an additional support for the stability of these
objects compared to their isothermal counterparts.  Indeed, these
results illustrate how the non-isothermal filaments could present
larger linear masses and shallower radial profiles than the
Ostriker-like filaments without being necessary unstable, where these
differences increase under the presence of large temperature
gradients.  

Available observations suggest rather shallow dust temperature
gradients.  For instance, the temperature gradient of the filament
B211/3 derived by \citep{palm12} indicates $A\simeq$ 0.022.  However,
dust and gas temperatures are likely not to be well coupled in
filaments.  This happens for densities larger than $\sim$ 3 $\cdot$
10$^4$ cm$^{-3}$ (Galli et al. 2002).  The derived central density in
B211/3 is $\sim$ 4.5 $\cdot$ 10$^4$ cm$^{-3}$ \citep[][see their
footnote 2]{palm12}, thus dust and gas temperatures are probably very
similar close to the axis.  They will decouple at some distance from
the axis, with cosmic rays ionization playing a role in heating up the
gas.  It is thus reasonable to expect gas temperature profiles steeper
than what dust measurements indicate.  This analytic work illustrates
how only dedicated and combined studies of both the mass distribution
and thermal structure within these objects (in addition to
simulations) can be then used to determine the physical state of
filaments in molecular clouds.

\begin{acknowledgements}
  This publication is supported by the Austrian Science Fund (FWF).
  Useful suggestions from an anonymous referee and from the Editor, M.
  Walmsley, are acknowledged.
\end{acknowledgements}

\begin{appendix}
\section{On the asymptotic slope of the density profile}
\label{sec:a1}
In this appendix we study the asymptotic behavior of the density
profile $\theta(x)$.  A similar problem has been previously discussed
in \citet{ostri64} and \citet{geh96} for filaments in equilibrium
ruled by different EOS.  In this case, we aim to investigate the
asymptotic behavior of the normalized radial profile ($\theta$) and
linear mass ($\Pi$) of a non-isothermal filament in equilibrium with a
temperature structure described by Eqs.~\ref{eq:tau1} or
\ref{eq:tau2}.

Let assume first a temperature profile like Eq.~\ref{eq:tau1}.
As explained in Sect.
\ref{sec:niso}, if we assume a power law dependence $\theta\sim
x^\alpha$, then $\alpha=\frac{d \ln \theta}{d \ln x}$.  
From Eq.~\ref{eq:tau1}, we transform Eq.~\ref{eq:ni_eq_norm} to calculate the quantity $\frac{d
  \ln \theta}{d \ln x}=\frac{x\theta^\prime}{\theta}$.  Assuming
moreover that $\theta(x)$ can be well approximated by a power law only
for $x>x^*$, then one obtains:

\begin{equation}
\frac{x\theta^\prime}{\theta}=-\frac{Kx}{1+Kx}-\frac{8}{1+Kx}\int_0^{x^*}
s\theta(s)ds-\frac{8}{1+Kx}\int_{x^*}^\infty
s^{\alpha+1}ds
\label{eq:alpha1}
\end{equation}
\noindent
For $x\rightarrow \infty$, the left hand side tends to $\alpha$, the
first term on the right hand side tends to $-1$, the second tends to 0,
and the third is proportional to $x^{\alpha-1}$ if $\alpha\neq -2$ and
to $\frac{\ln x}{x}$ if $\alpha=-2$.  We see now that $\alpha$ cannot
be larger than $-1$, otherwise the r.h.s. diverges.  It cannot be
smaller than (or equal to) $-2$, neither, since in this case $\alpha$
would tend to $-1$, contradicting the assumption that $\alpha\leq-2$.
A more careful analysis, based on the assumption that
$\theta(x)=x^\alpha+R(x)$ (where $R(x)$ is a small residual function)
leads indeed to the conclusion that $\alpha\in]-2,-1[$.  Consequently,
the integral $\Pi=\int 2\pi x \theta(x)dx$ is divergent.  

We employ now the temperature profile Eq. \ref{eq:tau2}. From it, Eq.
\ref{eq:ni_eq_norm} then becomes:
\begin{equation}
\frac{\theta^\prime}{\theta}=-\frac{K}{(1+x)[1+(K+1)x]}-
\frac{8(1+x)}{x[1+(K+1)x]}\int_0^x s \theta(s)ds.
\label{eq:thetadynamics}
\end{equation}
\noindent
Proceeding as before, it is easy to see that:
\begin{equation}
\alpha\simeq-\frac{8}{K+1}\displaystyle\lim_{x\to\infty}\int_0^x s\theta(s)ds.
\end{equation}
\noindent
Define now $I$ as $\int_0^{x^*}s\theta(s)ds$.  It is easy to obtain:
\begin{equation}
\alpha \simeq -\frac{8}{K+1}\cdot\displaystyle\lim_{x\to\infty} 
\begin{cases}
(I+\frac{x^{\alpha+2}-
(x^{*})^{\alpha+2}}{\alpha+2}) & \alpha\neq-2 
\\
(I+\ln x - \ln x^*) &\alpha=-2
\end{cases}
\label{eq:alpha2}
\end{equation}
\noindent
The case $\alpha=-2$ can be immediately ruled out because, for $x
\rightarrow \infty$ the r.h.s. of (\ref{eq:alpha2}) explodes to
$-\infty$, and analogously we cannot accept values of $\alpha$ larger
than $-2$, hence we conclude that $\alpha<-2$ necessarily.  This range
of values for $\alpha$ is particularly relevant in that it implies
finite masses, in fact the integral $\Pi=\int_0^\infty 2\pi x
\theta(x)dx$ does not diverge.
\end{appendix}

\end{document}